\documentclass{l4dc2027}

\title[GP-DDH]{Gaussian Process Data-Driven Hamiltonian\\for Probabilistic Safe Set Construction}
\usepackage{times}
\usepackage{xcolor}
\usepackage{graphicx}
\usepackage{booktabs}
\usepackage{placeins}
\usepackage{capt-of}

\newtheorem{assumption}{Assumption}

\newcommand{\avoidbrt}{\mathcal{S}_{\textit{Avoid}}}
\newcommand{\reachavoidbrt}{\mathcal{S}_{\textit{ReachAvoid}}}
\newcommand{\targetset}{\mathcal{X}_{\mathrm{T}}}
\newcommand{\unsafeset}{\mathcal{X}_{\mathrm{U}}}
\newcommand{\ddh}{\hat{H}}
\newcommand{\ddv}{\hat{V}}
\newcommand{\gph}{\hat{H}^{\beta}}
\newcommand{\gpv}{\hat{V}^{\beta}}

\newcommand{\betaq}{\beta_{\mathrm{Q}}}
\newcommand{\betabt}{\beta_{\mathrm{BT}}}

\author{%
    \Name{Seungseok Oh} \Email{seungseok.oh@snu.ac.kr}\\
    \Name{Junho Kim} \Email{junho9812@snu.ac.kr}\\
    \addr Seoul National University
    \AND
    \Name{Jason J. Choi} \Email{jjhchoi@ucla.edu}\\
    \addr University of California, Los Angeles
    \AND
    \Name{Namhoon Cho}\thanks{Corresponding author} \Email{namhoon.cho@snu.ac.kr}\\
    \addr Seoul National University
}

\begin{document}

\maketitle

\vspace{-2em}

\begin{abstract}%
    Hamilton--Jacobi (HJ) reachability is a general framework for verifying the safety of nonlinear systems, but it relies on an accurate dynamics model.
    The data-driven Hamiltonian (DDH) framework avoids system identification by lower-bounding the Hamiltonian directly from state--control--velocity data.
    However, it requires a Lipschitz bound on the unknown dynamics which is hard to certify from finite data, and as a single global constant this bound makes the computed safe sets conservative.
    We propose the Gaussian process data-driven Hamiltonian (GP-DDH), a fully data-driven framework that constructs safe sets with probabilistic guarantees under a GP prior on the dynamics.
    Furthermore, we suggest two methods to choose the confidence parameter $\beta$, both mathematically guaranteed to underbound the Hamiltonian.
    The computed safe set lies inside the true one with posterior probability at least $1-\Delta$ for any $\Delta\in(0,1)$.
    Numerical experiments show that GP-DDH recovers a bigger volume of the true safe set than DDH while still correctly providing a conservative approximation, even when DDH is given a Lipschitz constant computed from the true dynamics.
\end{abstract}


\section{Introduction}

Autonomous systems are moving into settings where a single safety violation can be catastrophic.
Many of them, such as aircraft in transition flight or legged robots on deformable terrain, have dynamics that are hard to identify.
Hamilton--Jacobi (HJ) reachability is rooted in model-based optimal control and provides a rigorous framework for verifying the safety of nonlinear systems~\citep{mitchell2005time,bansal2017hamilton}.
Without a model of the dynamics, however, it is of limited use for such systems.
This has motivated a growing interest in data-driven safety verification~\citep{wabersich2023data}.

The data-driven Hamiltonian (DDH) framework~\citep{choi2025data} removes the need for a model by lower-bounding the Hamiltonian directly with state--control--velocity observations, which yields an inner approximation of the true safe set.
Robust DDH extends it to measurement noise and bounded disturbances~\citep{bajelani2026robust}.
However, both require a valid Lipschitz or sensitivity bound on the dynamics, which is difficult to certify from finite data.
This bound is also a single global constant, so it inflates the uncertainty sets uniformly regardless of how the dynamics behave locally.

Gaussian process (GP) regression~\citep{williams2006gaussian} offers a local alternative: its posterior uncertainty is small near the observations and grows only where data are scarce.
GPs have been used to robustify control Lyapunov and barrier function constraints~\citep{castaneda2021gaussian,castaneda2021pointwise} and to enforce chance constraints on the barrier condition of control-affine systems whose drift and input gain are learned jointly by a matrix-variate GP~\citep{khojasteh2020probabilistic}.
Closest to our setting, \citet{akametalu2014reachability} and \citet{fisac2018general} augment a nominal model with a state-dependent disturbance whose bounds are learned by a GP and compute the safe set from a Hamilton--Jacobi--Isaacs (HJI) equation in which this disturbance acts adversarially.
Unlike these methods, we require no nominal model and allow the uncertainty to depend on the input.

In this work, we propose the \emph{Gaussian process data-driven Hamiltonian} (GP-DDH), a fully data-driven framework that constructs safe sets with probabilistic guarantees.
Using a vector-valued GP prior on the dynamics, it lower-bounds the Hamiltonian with the posterior mean and the posterior uncertainty scaled by a confidence parameter $\beta$.
Specifically, our contributions are as follows:

\begin{itemize}
    \item We prove that the GP-DDH safe sets for the \textit{Avoid} and \textit{ReachAvoid} problems lie inside the true safe sets with posterior probability at least $1-\Delta$ (Sections~\ref{sec:prob_safety} and~\ref{sec:borell_tis}).
    \item We give two ways to choose $\beta$, each with its own probabilistic guarantee. The Borell--TIS choice holds uniformly over the continuous state--control domain (Section~\ref{sec:borell_tis}). The order-statistic choice has exact coverage $1-\Delta$ at the state--control grids used by a numerical HJ solver and gives larger safe sets (Section~\ref{sec:order_statistic}).
    \item We test GP-DDH on dynamics sampled from a GP prior and on a random polynomial benchmark (Section~\ref{sec:experiments}). GP-DDH recovers more of the true safe set than DDH, even when DDH uses a Lipschitz constant computed from the true dynamics.
\end{itemize}

\section{Problem Formulation and Background}
\label{sec:problem}

\subsection{Problem Formulation}

We consider the nonlinear control system
\begin{equation}
    \label{eq:dynamics}
    \dot{x}(s) = f \bigl(x(s), u(s)\bigr), \qquad s \in [-t, 0], \quad x(-t) = x,
\end{equation}
where $x(s)\in\mathcal X\subseteq\mathbb R^{n_x}$, $u(s)\in\mathcal U\subseteq\mathbb R^{n_u}$, and $t\in[0,T]$.
An admissible control signal $u(\cdot)$ is a measurable function taking values in the compact set $\mathcal U$.
The dynamics $f:\mathcal X\times\mathcal U\to\mathbb R^{n_x}$ are continuous and Lipschitz continuous in $x$ uniformly over $u\in\mathcal U$, so each admissible control signal generates a unique trajectory from every initial state.
Both $f$ and its Lipschitz constant are unknown, and only a finite set of noise-free observations is given:
\begin{equation}
    \label{eq:dataset}
    \mathcal D = \{(x^i, u^i, v^i)\}_{i = 1}^{N}, \qquad v^i = f(x^i, u^i).
\end{equation}

The unsafe set and an optional target set are represented by Lipschitz-continuous functions $g,l:\mathcal X\to\mathbb R$ as $\unsafeset=\{x:g(x)<0\}$ and $\targetset=\{x:l(x)\ge0\}$.
The \textit{Avoid} backward reachable tube (BRT) is the set of initial states from which every admissible control signal drives the trajectory into $\unsafeset$ at some $s\in[-t,0]$.
The \textit{ReachAvoid} BRT is the set of initial states from which some admissible control signal drives the trajectory into $\targetset$ at some $\tau\in[-t,0]$ without entering $\unsafeset$ on $[-t,\tau]$.
We refer to the complement of the \textit{Avoid} BRT and to the \textit{ReachAvoid} BRT as the safe sets, and denote them by $\avoidbrt(t)$ and $\reachavoidbrt(t)$, respectively.

\paragraph{Goal.} Given $\Delta\in(0,1)$, we seek sets computed from $\mathcal D$ alone that lie inside the safe sets with probability at least $1-\Delta$ under a Gaussian process model of $f$.

\subsection{Hamilton--Jacobi Reachability}

HJ reachability~\citep{mitchell2005time,bansal2017hamilton} encodes each safe set as the zero-superlevel set of a value function computed by dynamic programming.
For the \textit{Avoid} problem, the value function $V(x,t):=\sup_{u(\cdot)}\min_{s\in[-t,0]}g(x(s))$ satisfies $\avoidbrt(t)=\{x:V(x,t)\ge0\}$ and is the viscosity solution~\citep{bardi1997optimal} of the HJ variational inequality (HJ-VI)~\citep{fisac2015reach}
\begin{equation}
    \label{eq:hjvi_avoid}
    0 = \min \bigl\{ g(x) - V,\, -D_t V + H(x, D_x V) \bigr\}, \qquad V(x, 0) = g(x).
\end{equation}
For the \textit{ReachAvoid} problem, we overload the notation and define $V(x,t):=\sup_{u(\cdot)}\max_{s\in[-t,0]}\allowbreak\min\{l(x(s)),\allowbreak\min_{\tau\in[-t,s]}g(x(\tau))\}$,
which satisfies $\reachavoidbrt(t)=\{x:V(x,t)\ge0\}$ and is the viscosity solution of
\begin{equation}
    \label{eq:hjvi_reachavoid}
    0 = \min \bigl\{ g(x) - V,\, \max \{ l(x) - V,\, -D_t V + H(x, D_x V) \} \bigr\}, \qquad V(x, 0) = \min \{ l(x), g(x) \}.
\end{equation}
In both HJ-VIs, the dynamics enter only through the Hamiltonian
\begin{equation}
    \label{eq:hamiltonian}
    H(x, p) := \max_{u \in \mathcal U} p^\top f(x, u),
\end{equation}
where $p=D_xV(x,t)$ is the costate.
We refer to the safe sets computed with this exact Hamiltonian as the true safe sets.

\subsection{Data-Driven Hamiltonian}
\label{sec:ddh}

Since the dynamics enter the HJ-VIs only through $H$, the DDH framework~\citep{choi2025data} replaces $H$ with a lower bound constructed directly from $\mathcal D$ instead of identifying $f$.

\begin{definition}[Data-driven Hamiltonian]
    \label{def:ddh}
    \begin{equation}
        \ddh(x, p) := \max_{i \in \{1, \dots, N\}} \min_{\hat v \in v^i \oplus \mathcal E(x;x^i)} p^\top \hat v,
    \end{equation}
    where $\mathcal E(x;x^i)\subset\mathbb R^{n_x}$ is an uncertainty set containing $f(x,u^i)-v^i$.
\end{definition}
For instance, a Lipschitz constant $L_x$ with $\|f(x,u)-f(x',u)\|_2\le L_x\|x-x'\|_2$ for all $u\in\mathcal U$ yields the ball $\mathcal E(x;x^i)=\{e:\|e\|_2\le L_x\|x-x^i\|_2\}$, for which the inner minimum equals $p^\top v^i-L_x\|x-x^i\|_2\|p\|_2$.
Because each $\mathcal E(x;x^i)$ contains $f(x,u^i)-v^i$ and each observed input is admissible, $\ddh(x,p)\le H(x,p)$ for all $(x,p)$.

Let $\ddv(x,t)$ denote the viscosity solution of \eqref{eq:hjvi_avoid} or \eqref{eq:hjvi_reachavoid} with $H$ replaced by $\ddh$ and the same terminal condition.
The following theorem states that DDH is conservative.

\begin{theorem}[{\citealt[Theorem~1]{choi2025data}}]
    \label{thm:ddh_safety}
    For the \textit{Avoid} and \textit{ReachAvoid} problems,
    \begin{equation}
    \ddv(x, t) \le V(x, t)
    \end{equation}
    on $\mathcal X \times [0, T]$; hence $\{x:\ddv(x,t)\ge0\}$ is an inner approximation of the corresponding safe set.
\end{theorem}

\section{Gaussian Process Data-Driven Hamiltonian}
\label{sec:gpddh}

\subsection{Gaussian Process Model of the Dynamics}
\label{sec:gp_regression}

We start by assuming that the unknown dynamics follow a vector-valued GP prior, which is a commonly used nonparametric model for unknown functions~\citep{williams2006gaussian}.

\begin{assumption}[Vector-valued GP prior]
    \label{asm:vector_valued_gp}
    The set $\mathcal X\times\mathcal U$ is compact, and the unknown dynamics follow the separable vector-valued GP prior
    \begin{equation}
        f \sim \mathcal{GP} \bigl(m_0, \,k(\cdot, \cdot) B \bigr),
    \end{equation}
    where $m_0:\mathcal X\times\mathcal U\to\mathbb R^{n_x}$ is a continuous prior mean, $k$ is a scalar covariance kernel over state--control pairs, and $B\in\mathbb R^{n_x\times n_x}$ is a symmetric positive-definite matrix that specifies covariance across the components of $f$.
\end{assumption}

This separable structure is the standard way to share a kernel across the outputs of a vector-valued GP~\citep{alvarez2012kernels}.
The prior mean $m_0$ can encode a nominal model when one is available.
The kernel hyperparameters and $B$ are treated as fixed, for example after fitting them to $\mathcal D$ by maximizing the marginal likelihood.

If $m_0$ is smooth, the sample paths of $f$ are continuously differentiable with probability one for the RBF, Mat\'ern-5/2, and linear kernels used in Section~\ref{sec:experiments}.
Hence $f$ is Lipschitz continuous on the compact set $\mathcal X\times\mathcal U$, as required in Section~\ref{sec:problem}.

We write $z:=(x,u)$ and $z_i:=(x^i,u^i)$, and let $K\in\mathbb R^{N\times N}$ with $K_{ij}:=k(z_i,z_j)$, $\mathbf k(z):=[k(z,z_1),\ldots,k(z,z_N)]^\top$, and $Y,M_0\in\mathbb R^{N\times n_x}$ stack $v^i$ and $m_0(z_i)$, respectively.
Assuming that $K$ is positive definite (e.g., distinct observed inputs and a strictly positive-definite $k$; see Appendix~\ref{app:details} for the numerical jitter), noise-free conditioning~\citep{williams2006gaussian} gives the posterior $f|\mathcal D\sim\mathcal{GP}(\mu_0,\Sigma_0)$ with
\begin{equation}
    \label{eq:gp_posterior}
    \mu_0(z) = m_0(z) + (Y - M_0)^\top K^{-1} \mathbf k(z), \qquad
    \Sigma_0(z, z') = \bigl(k(z, z') - \mathbf k(z)^\top K^{-1} \mathbf k(z') \bigr) B,
\end{equation}
and we write $\Sigma_0(z):=\Sigma_0(z,z)$.

\subsection{Definition}
\label{sec:gpddh_def}

Since the GP posterior describes the uncertainty in the dynamics at unobserved states and controls, we define a local uncertainty set $\mathcal F^{\beta}(z)$ in the shape of an ellipsoid with a confidence parameter $\beta>0$:
\begin{equation}
    \label{eq:gp_ellipsoid}
    \mathcal F^{\beta}(z) := \bigl\{ v\in\mathbb R^{n_x} : \bigl(v-\mu_0(z)\bigr)^\top \overline \Sigma(z)^{-1} \bigl(v-\mu_0(z)\bigr) \le \beta^2 \bigr\},
\end{equation}
where $\overline\Sigma(z):=\Sigma_0(z)+\eta^2I_{n_x}$ with $\eta>0$ is the regularized covariance.
Because $\Sigma_0(z)$ vanishes at the observed inputs, the regularization keeps $\overline\Sigma(z)$ invertible.
With this uncertainty set, we define the GP-DDH as follows.

\begin{definition}[Gaussian process data-driven Hamiltonian]
    \label{def:gpddh}
    \begin{equation}
        \label{eq:gph}
        \gph(x, p) := \max_{u \in \mathcal U} \min_{v \in \mathcal F^{\beta}(x, u)} p^\top v
        = \max_{u \in \mathcal U} \Bigl\{ p^\top \mu_0(x, u) - \beta \sqrt{p^\top \overline \Sigma(x, u) p} \Bigr\},
    \end{equation}
    where the equality follows from the closed-form minimum of a linear function over an ellipsoid.
\end{definition}

\begin{figure}[t]
    \centering
    \includegraphics[width=0.80\linewidth]{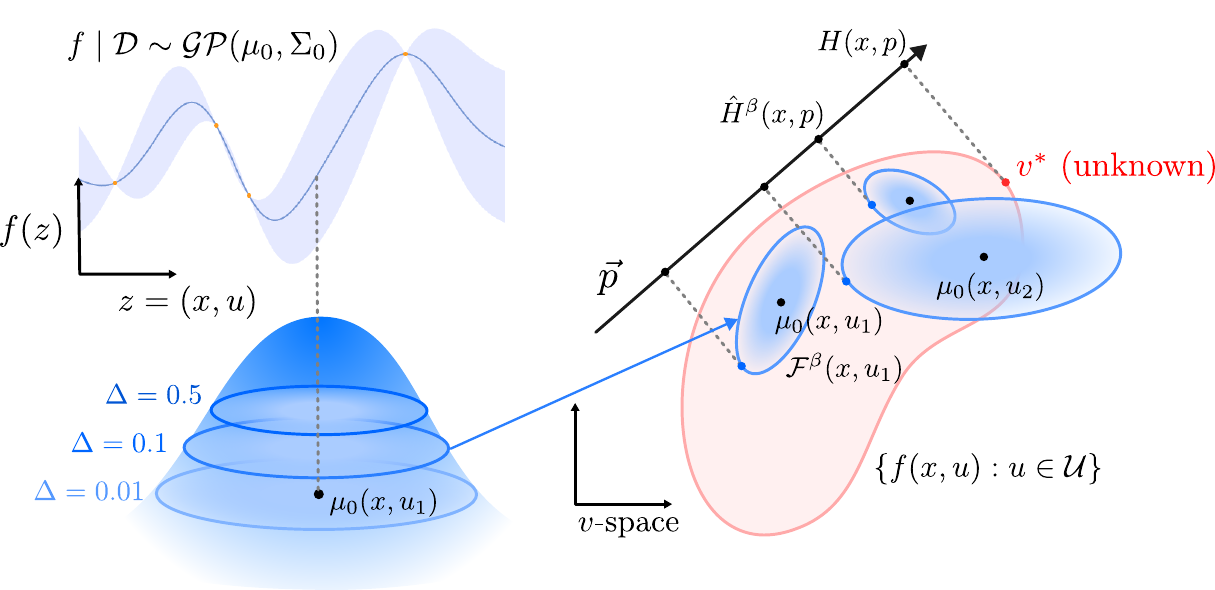}
    \caption{GP-DDH at a fixed state $x$ and costate $p$. Left: posterior of $f$ (one output component) and posterior density of $f(x,u_1)$, whose level sets are the ellipsoids $\mathcal F^\beta(x,u_1)$; a smaller $\Delta$ gives a larger $\beta$. Right: the worst case over each ellipsoid along $p$ (blue dots), maximized over $u$, gives $\gph(x,p)\le H(x,p)=p^\top v^*$ whenever every ellipsoid contains the true velocity.}
    \label{fig:gpddh_concept}
    \vspace{-2em}
\end{figure}

The schematic of GP-DDH is shown in Figure~\ref{fig:gpddh_concept}.
Unlike DDH, which maximizes only over the observed inputs $u^i$, GP-DDH maximizes over all admissible inputs $u\in\mathcal U$ because $\mathcal F^\beta$ is defined at every state--control pair.

The confidence parameter $\beta$ determines the conservativeness of GP-DDH.
A larger $\beta$ gives a larger ellipsoid $\mathcal F^\beta(z)$, which lowers $\gph$ and makes GP-DDH more conservative.
In Section~\ref{sec:choosing}, we give two ways to choose $\beta$ as a function of $\Delta\in(0,1)$ that yield a probabilistic safety guarantee.

We also define the corresponding value function $\gpv(x,t)$ as the viscosity solution of \eqref{eq:hjvi_avoid} or \eqref{eq:hjvi_reachavoid} with $H$ replaced by $\gph$ and the same terminal condition.

\subsection{Conservativeness of GP-DDH}
\label{sec:prob_safety}

GP-DDH yields an inner approximation of the true safe set when $\beta$ is large enough, and we now make this condition precise.

As for DDH in Theorem~\ref{thm:ddh_safety}, it suffices that $\gph(x,p)\le H(x,p)$ for all $(x,p)$, and this holds whenever $f(z)\in\mathcal F^\beta(z)$ for all $z=(x,u)$.
To write this condition as a single inequality on $\beta$, we measure the distance from $\mu_0(z)$ to $f(z)$ in the metric of the ellipsoid.

For $(z,p)\in\mathcal I:=(\mathcal X\times\mathcal U)\times\mathbb S^{n_x-1}$, where $\mathbb S^{n_x-1}$ is the unit sphere, we define the normalized posterior residual $W(z,p)$ and, by the Cauchy--Schwarz inequality, its supremum over costate directions:
\begin{equation}
    \label{eq:W}
    W(z, p) := \frac{p^\top \bigl( \mu_0(z) - f(z) \bigr)} { \sqrt{p^\top \overline \Sigma(z) p}}, \qquad
    d(z) := \sup_{p \in \mathbb S^{n_x-1}} W(z, p) = \Bigl\| \overline \Sigma(z)^{-1/2} \bigl( \mu_0(z) - f(z) \bigr) \Bigr\|_2 .
\end{equation}
Thus $d(z)\le\beta$ if and only if $f(z)\in\mathcal F^\beta(z)$.
Let
\begin{equation}
    \label{eq:M}
    M(f):=\sup_{z\in\mathcal X\times\mathcal U}d(z)=\sup_{(z,p)\in\mathcal I}W(z,p)
\end{equation}
be the largest normalized residual over the domain. Then, it suffices that $\beta\ge M(f)$ for GP-DDH to be conservative, as Lemma~\ref{lem:conservative} shows.

\begin{lemma}[Conservative safe sets]
    \label{lem:conservative}
    Suppose that the HJ-VIs \eqref{eq:hjvi_avoid} and \eqref{eq:hjvi_reachavoid} satisfy the comparison principle with common terminal and boundary data.
    If $M(f)\le\beta$, then $\gph(x,p)\le H(x,p)$ for all $x\in\mathcal X$ and $p\in\mathbb R^{n_x}$, and $\gpv(x,t)\le V(x,t)$ on $\mathcal X\times[0,T]$.
    Consequently, $\{x: \gpv(x, t) \ge 0\} \subseteq \{x:V(x, t) \ge 0\}$ for all $t \in [0, T]$.
\end{lemma}

\begin{proof}
    If $M(f)\le\beta$, then $f(z)\in\mathcal F^\beta(z)$ for every $z$, so the inner minimum in \eqref{eq:gph} is at most $p^\top f(x,u)$.
    Taking the maximum over $u$ gives $\gph\le H$, and the comparison principle yields $\gpv\le V$ as in Theorem~\ref{thm:ddh_safety}.
\end{proof}

Thus $\beta=M(f)$ is the smallest value for which Lemma~\ref{lem:conservative} guarantees safety, and any larger $\beta$ only adds conservatism.
However, $M(f)$ cannot be computed because $f$ is unknown --- our knowledge of $f$ given $\mathcal D$ is the GP posterior, which induces a distribution of $M(f)$ under Assumption~\ref{asm:vector_valued_gp}.
It therefore remains to choose $\beta$ that upper-bounds $M(f)$ with a prescribed posterior probability, which is the subject of Section~\ref{sec:choosing}.

\section{Choosing the Confidence Parameter \texorpdfstring{$\beta$}{beta} with Probabilistic Guarantees}
\label{sec:choosing}

The smallest $\beta$ that upper-bounds $M(f)$ with posterior probability $1-\Delta$ is the $(1-\Delta)$-quantile of $M(f)$ given $\mathcal D$.
This quantile has no closed form, however, so we give two ways to choose $\beta$ instead, each with its own probabilistic guarantee.
The first is an analytic upper bound on the quantile from the Borell--TIS inequality (Section~\ref{sec:borell_tis}).
The second is a practical estimate on the solver grid from an order statistic of posterior sample paths (Section~\ref{sec:order_statistic}).

\subsection{Analytic Confidence Parameter from the Borell--TIS Inequality}
\label{sec:borell_tis}

We define the Borell--TIS confidence parameter $\betabt(\Delta)$ as
\begin{equation}
    \label{eq:beta_bt_def}
    \boxed{\betabt(\Delta) := \mathbb E[M(f)\bigm|\mathcal D] + \sqrt{2\log(1/\Delta)}.}
\end{equation}
Its dependence on $\Delta$ is explicit: it grows like $\sqrt{2\log(1/\Delta)}$, while the data and the kernel enter only through the expectation of $M(f)$.
We now show that $\betabt$ gives a probabilistic guarantee.

\begin{theorem}[Probabilistic safety]
    \label{thm:prob_safety}
    Suppose that Assumption~\ref{asm:vector_valued_gp} holds and that the HJ-VIs \eqref{eq:hjvi_avoid} and \eqref{eq:hjvi_reachavoid} satisfy the comparison principle with common terminal and boundary data.
    If we set $\beta = \betabt(\Delta)$ for $\Delta\in(0,1)$,
    \begin{equation}
        \label{eq:beta_exact}
        \mathbb P \bigl( M(f) \le \beta \bigm| \mathcal D \bigr) \ge 1 - \Delta,
    \end{equation}
    and consequently
    \begin{equation}
        \label{eq:prob_safety}
        \mathbb P \bigl( \gpv(x, t) \le V(x, t) \ \ \forall (x, t) \in \mathcal X \times [0, T] \bigm| \mathcal D \bigr) \ge 1 - \Delta.
    \end{equation}
    With the same probability, $\{x: \gpv(x, t) \ge 0\} \subseteq \{x:V(x, t) \ge 0\}$ for all $t \in [0, T]$.
\end{theorem}

\begingroup
\renewcommand{\proofname}{Sketch of proof}
\begin{proof}
    Given $\mathcal D$, $W$ is a centered Gaussian random field with variance at most one, so the Borell--TIS inequality~\citep[Theorem~2.1.1]{adler2007random} bounds the upper tail of its supremum $M(f)$ and gives \eqref{eq:beta_exact}.
    Once \eqref{eq:beta_exact} holds, \eqref{eq:prob_safety} and the safe-set inclusion follow from Lemma~\ref{lem:conservative}.
    Appendix~\ref{app:proofs} gives the full proof.
\end{proof}
\endgroup

The guarantee in Theorem~\ref{thm:prob_safety} holds uniformly over the continuous domain $\mathcal X\times\mathcal U$.
In practice, the expectation $\mathbb E[M(f)|\mathcal D]$ can be approximated from posterior sample paths, as detailed in Appendix~\ref{app:details}.

\subsection{Practical Confidence Parameter from an Order Statistic}
\label{sec:order_statistic}

In practice, most numerical HJ PDE solvers evaluate the Hamiltonian only on a finite grid of states and controls.
It can therefore be sufficient to guarantee $f(z)\in\mathcal F^\beta(z)$ only on this grid, which leads to a different and less conservative choice of $\beta$.

Let $\mathcal X_h\subset\mathcal X$ and $\mathcal U_h\subset\mathcal U$ be the state and control grids of the solver, where the subscript $h$ marks quantities defined on the grid, let $\gph_h$ denote \eqref{eq:gph} with the maximum taken over $\mathcal U_h$, and let $\mathcal Z_h\subseteq\mathcal X\times\mathcal U$ be a finite calibration set, typically $\mathcal Z_h=\mathcal X_h\times\mathcal U_h$, with $M_h(f):=\max_{z\in\mathcal Z_h}d(z)$.
For a given numerical scheme, let $\gpv_h$ and $V_h$ denote its solutions on $\mathcal X_h$ with $\gph_h$ and $H$ as the Hamiltonian, interpolated linearly in time between the time steps.

To estimate the $(1-\Delta)$-quantile of $M_h(f)$ given $\mathcal D$ directly from posterior sample paths, we draw $S$ sample paths $f^{(1)},\ldots,f^{(S)}$ i.i.d. from the joint posterior of $\{f(z):z\in\mathcal Z_h\}$ and compute $M_{h,s}:=M_h(f^{(s)})$.
Then we define
\begin{equation}
    \label{eq:beta_quantile}
    \boxed{\betaq(\Delta) := M_{h,(k)}, \qquad k = \bigl\lceil (S+1)(1-\Delta) \bigr\rceil,}
\end{equation}
where $M_{h,(k)}$ is the $k$-th smallest of the $M_{h,s}$ and $S\ge1/\Delta-1$ ensures $k\le S$.

\begin{theorem}[Probabilistic safety on the grid]
    \label{thm:quantile}
    Suppose that Assumption~\ref{asm:vector_valued_gp} holds and that $\mathcal Z_h\supseteq\mathcal X_h\times\mathcal U_h$.
    If we set $\beta=\betaq(\Delta)$ for $\Delta\in(0,1)$, 
    \begin{equation}
        \label{eq:quantile_event}
        \mathbb P \bigl( M_h(f) \le \beta \bigm| \mathcal D \bigr) \ge 1 - \Delta,
    \end{equation}
    and consequently
    \begin{equation}
        \label{eq:prob_safety_grid}
        \mathbb P \bigl( \gpv_h(x, t) \le V_h(x, t) \ \ \forall (x, t) \in \mathcal X_h \times [0, T] \bigm| \mathcal D \bigr) \ge 1 - \Delta,
    \end{equation}
    where both probabilities are taken jointly over the posterior of $f$ and the sample paths.
    With the same probability, $\{x\in\mathcal X_h: \gpv_h(x, t) \ge 0\} \subseteq \{x\in\mathcal X_h:V_h(x, t) \ge 0\}$ for all $t \in [0, T]$.
\end{theorem}

The proof of Theorem~\ref{thm:quantile}, given in Appendix~\ref{app:proofs}, rests on the exchangeability argument behind Monte Carlo tests~\citep{dwass1957modified,barnard1963discussion} and split conformal prediction~\citep{vovk2005algorithmic,lei2018distribution}.
Here the calibration samples are posterior draws rather than held-out data, so the guarantee is Bayesian, as in Theorem~\ref{thm:prob_safety}.


\subsection{Choosing between \texorpdfstring{$\betabt$}{beta\_BT} and \texorpdfstring{$\betaq$}{beta\_Q}}
\label{sec:choice}

The parameter $\betabt$ follows the Borell--TIS bound, which holds uniformly over the continuous domain $\mathcal X\times\mathcal U$.
It is typically larger than $\betaq$, so its safe sets are more conservative (Section~\ref{sec:experiments}).
The parameter $\betaq$ gives less conservative safe sets, but its guarantee holds only on the solver grid $\mathcal X_h$.
Choosing between them is therefore a design decision between a guarantee over the continuous domain and a larger safe set.

\section{Numerical Experiments}
\label{sec:experiments}

We evaluate GP-DDH on two examples.
The first example takes a single sample path of the GP prior as the true dynamics.
Because Assumption~\ref{asm:vector_valued_gp} holds by construction for this system, we can check whether the probabilistic guarantee is valid.
Secondly, as a more practical example, we adopt the random polynomial benchmark of \citet{choi2025data} for which the GP model is misspecified.

In both examples, we assume that the true dynamics are unknown and compare the safe sets of GP-DDH and DDH computed from the same noise-free dataset.
GP-DDH uses the separable model of Assumption~\ref{asm:vector_valued_gp} with a full matrix $B$ and $m_0\equiv0$.
DDH uses a Lipschitz constant $L_x$ computed from the true dynamics, which favors DDH because such a constant is not available from data alone.
The setups and numerical details are given in Appendices~\ref{app:details}--\ref{app:poly_details}.

\subsection{GP-Sampled Dynamics}
\label{sec:gp_running_example}

We consider a system whose true dynamics are a single sample path
\begin{equation}
    \label{eq:gp_sampled_oscillator}
    f_{\mathrm{true}} \sim \mathcal{GP}(0, k_{\mathrm{RBF}} B), \qquad B = 0.15^2\begin{bmatrix}1 & 0.6\\0.6 & 0.8\end{bmatrix},
\end{equation}
with $x\in\mathbb R^2$ and $u\in\mathbb R$.
We solve an \textit{Avoid} problem for $T=1$ with the unsafe set $\{x:|x_1|>1\}$.
From $N=100$ noise-free observations of $f_{\mathrm{true}}$, the GP-DDH safe sets are computed for several values of $\Delta$ and compared with the true safe set and the DDH safe set.

\paragraph{Confidence parameters.}

\begin{table}[!htbp]
    \centering
    \small
    \caption{Results for the GP-sampled \textit{Avoid} problem.
    The empirical probability is the mean ($\pm$ standard deviation) of $\mathbb P(M_h(f)\le\beta)$ over $100$ repetitions, each of which recomputes $\beta$ from new sample paths.
    Recovery is the percentage of the safe states recovered from the true safe set.}
    \label{tab:gp_sampled_results}
    \begin{tabular}{rrrrrrrr}
        \toprule
        & \multicolumn{2}{c}{$\beta$} & \multicolumn{3}{c}{Empirical $\mathbb P(M_h(f)\le\beta)$} & \multicolumn{2}{c}{Recovery (\%)} \\
        \cmidrule(lr){2-3} \cmidrule(lr){4-6} \cmidrule(lr){7-8}
        $\Delta$ & $\betabt$ & $\betaq$ & \makebox[1.25cm][c]{Target} & \makebox[1.9cm][c]{$\betabt$} & \makebox[1.9cm][c]{$\betaq$} & $\betabt$ & $\betaq$ \\
        \midrule
        0.5 & 5.48 & 4.26 & 0.5 & $0.997(\pm0.000)$ & $0.500(\pm0.016)$ & 69.8 & 78.8 \\
        0.1 & 6.45 & 4.73 & 0.9 & $1.000(\pm0.000)$ & $0.900(\pm0.010)$ & 63.0 & 74.6 \\
        0.01 & 7.33 & 5.32 & 0.99 & $1.000(\pm0.000)$ & $0.990(\pm0.003)$ & 54.5 & 70.9 \\
        0.001 & 8.02 & 5.63 & 0.999 & $1.000(\pm0.000)$ & $0.999(\pm0.001)$ & 48.7 & 68.3 \\
        \bottomrule
    \end{tabular}
\end{table}

Table~\ref{tab:gp_sampled_results} lists $\betabt(\Delta)$ and $\betaq(\Delta)$ for each $\Delta$, and checks the probability in \eqref{eq:quantile_event} by the fraction of safe functions to $10{,}000$ possible functions.
Each probability is computed over $100$ independent repetitions.

\begin{figure*}[!htbp]
    \centering
    \begin{minipage}[t]{0.48\textwidth}
        \centering
        \includegraphics[width=1.00\linewidth]{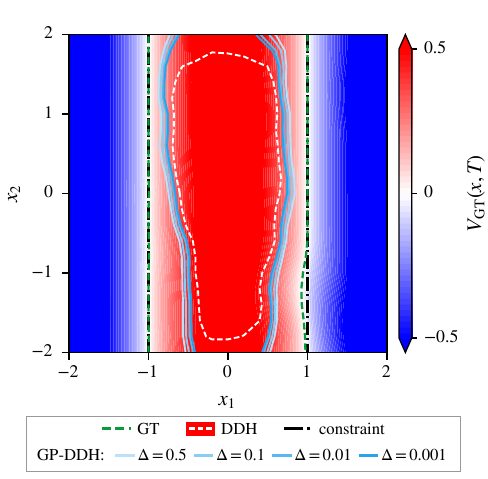}
        \caption{\textit{Avoid} safe sets of GP sampled dynamics at $T=1$ over the true value function. The GP-DDH sets use $\betaq$, are nested in $\Delta$, and contain the DDH set.}
        \label{fig:gp_safe_sets}
    \end{minipage}\hfill
    \begin{minipage}[t]{0.48\textwidth}
        \centering
        \includegraphics[width=1.00\linewidth]{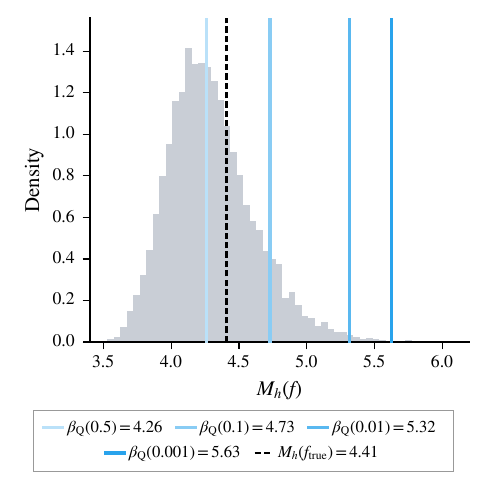}
        \caption{Histogram of $M_h(f)$ for $10{,}000$ posterior sample paths $f$, with $\betaq$ and $M_h(f_{\mathrm{true}})$ marked.}
        \label{fig:gp_sampled_supM_distribution}
    \end{minipage}
    \vspace{-2em}
\end{figure*}

The mean empirical probability for $\betabt$ is $0.997$ even for $\Delta=0.5$, far above the target $1-\Delta$.
The excess comes from the tail term $\sqrt{2\log(1/\Delta)}$, which bounds the fluctuation of $M_h(f)$ by that of a standard Gaussian variable, whereas $M_h(f)$ in Figure~\ref{fig:gp_sampled_supM_distribution} has a standard deviation of $0.32$.
The mean empirical probability for $\betaq$ lies within $0.001$ of the target $1-\Delta$, as Theorem~\ref{thm:quantile} predicts.
$\betaq$ is smaller than $\betabt$ for every $\Delta$, which makes its safe sets less conservative.

Both confidence parameters therefore reach the target probability $1-\Delta$ on average.
This supports Theorems~\ref{thm:prob_safety} and~\ref{thm:quantile} empirically under Assumption~\ref{asm:vector_valued_gp}.

\paragraph{Safe sets.}
Figure~\ref{fig:gp_safe_sets} shows that the GP-DDH sets shrink in a nested manner as $\Delta$ decreases, because a smaller $\Delta$ gives a larger $\betaq(\Delta)$ and hence larger ellipsoids $\mathcal F^{\betaq}(z)$.
Even the largest of these sets, obtained for $\Delta=0.5$, contains no unsafe state, although $M_h(f_{\mathrm{true}})=4.41$ exceeds the corresponding $\betaq=4.26$.
This is because the event in \eqref{eq:quantile_event} is sufficient but not necessary for the inclusion of the safe sets.
Figure~\ref{fig:gp_sampled_supM_distribution} places $M_h(f_{\mathrm{true}})$ at the $68$th percentile, so the displayed $f_{\mathrm{true}}$ is not a favorable draw.

Every GP-DDH set with $\betaq$ contains the DDH set and recovers between $68.3\%$ and $78.8\%$ of the true safe set, whereas DDH recovers $51.3\%$.
The advantage arises because each ellipsoid $\mathcal F^{\betaq}(z)$ follows the local posterior uncertainty, whereas DDH uses a single global Lipschitz constant.

\subsection{Random Polynomial System}
\label{sec:2d_rand_poly}

We consider a random polynomial system with dynamics given by
\begin{equation}
    \label{eq:2d_poly_dynamics}
    f_i(x, u) = p_i(u_1, u_2) + q_i(u_1, u_2) x_1 + r_i(u_1, u_2) x_2, \quad i \in \{1, 2\},
\end{equation}
where $p_i$, $q_i$, and $r_i$ are quadratic polynomials with randomly drawn coefficients.
We solve a \textit{ReachAvoid} problem that reaches a circle around the origin while staying inside the unit square, from $N=50$ observations in that square.
We compare three kernels: RBF, RBF plus a linear kernel (RBF+Lin), and Mat\'ern-5/2.
The GP model is fitted to the dataset by maximizing the marginal likelihood over the hyperparameters and $B$.


\begin{figure}[t]
    \centering
    \includegraphics[width=0.92\linewidth]{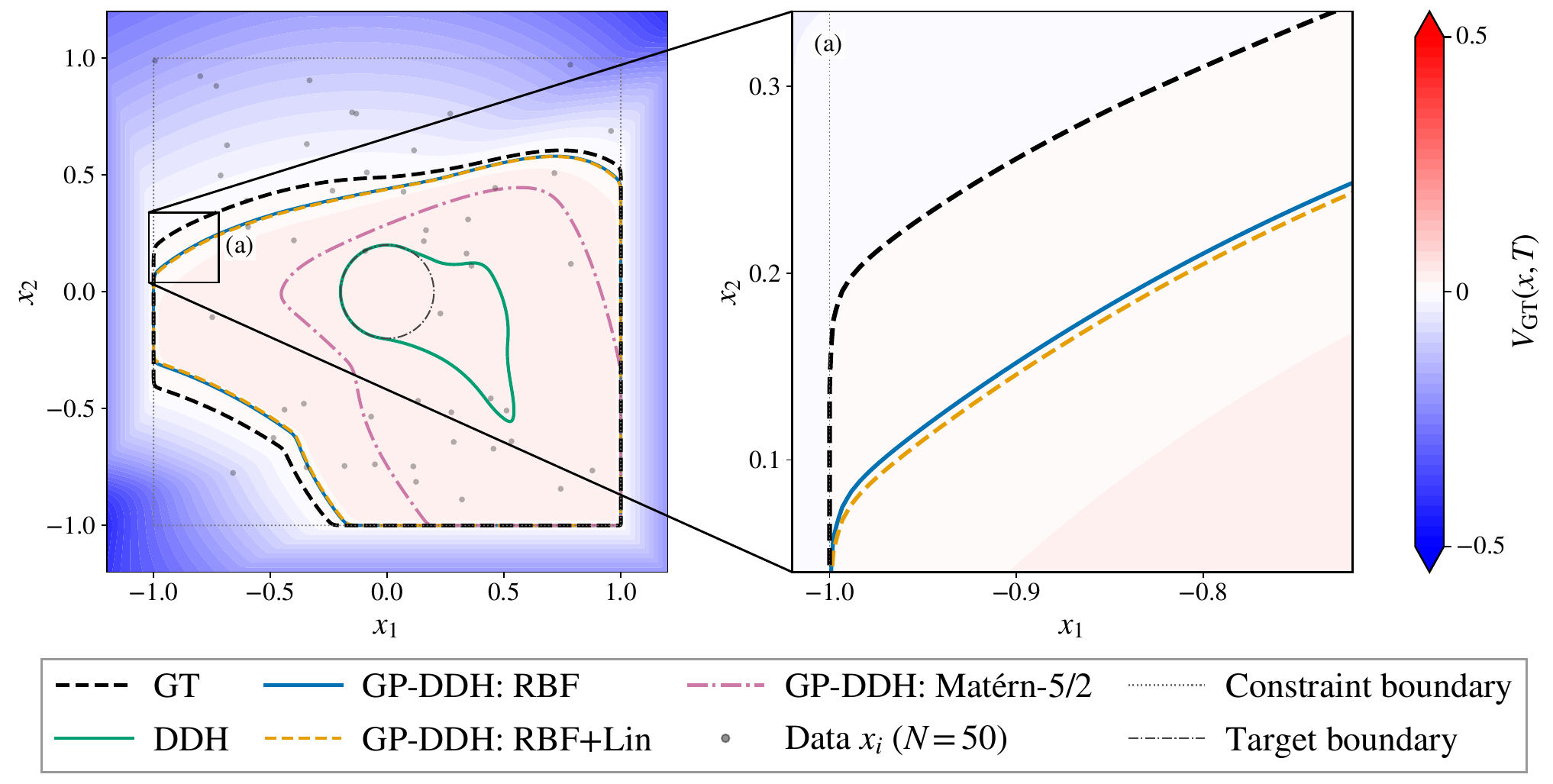}
    \caption{\textit{ReachAvoid} safe sets of the random polynomial system at $T=0.5$, with GP-DDH using $\betaq$ for $\Delta=0.1$. Left: full domain with the $N=50$ data states. Right: close-up of (a).}
    \label{fig:poly2d_comparison}
    \vspace{-0.6em}
    \small
    \captionof{table}{Confidence parameters and safe set results from the random polynomial system.}
    \vspace{0.5em}
    \label{tab:poly2d_results}
    \begin{tabular}{lrrrr}
    \toprule
    Method & $\betabt$ & $\betaq$ & $M_h(f_{\mathrm{true}})$ & Recovery (\%) \\
    \midrule
    DDH ($L_x=6.63$) & -- & -- & -- & 11.4 \\
    GP-DDH (RBF) & 5.55 & 4.11 & 3.28 & 92.4 \\
    GP-DDH (RBF+Lin) & 5.59 & 4.18 & 3.32 & 91.9 \\
    GP-DDH (Mat\'ern-5/2) & 6.03 & 4.44 & 4.30 & 57.0 \\
    \bottomrule
    \end{tabular}
\end{figure}

\paragraph{Confidence parameters.}
Table~\ref{tab:poly2d_results} lists $\betabt(\Delta)$ and $\betaq(\Delta)$ for $\Delta=0.1$ with each kernel.
For all three kernels, $M_h(f_{\mathrm{true}})$ lies below both confidence parameters, so the event in \eqref{eq:quantile_event} holds for $f_{\mathrm{true}}$ on $\mathcal Z_h$ even though the GP model is misspecified.
Every GP-DDH set therefore lies inside the true safe set.

\paragraph{Safe sets.}
Figure~\ref{fig:poly2d_comparison} shows that GP-DDH recovers $92.4\%$ and $91.9\%$ of the true safe set with RBF and RBF+Lin, while DDH recovers only $11.4\%$.

The performance of GP-DDH depends on how well the kernel models the true dynamics.
Mat\'ern-5/2 has a larger posterior uncertainty because of its rougher sample paths, and hence larger ellipsoids $\mathcal F^{\betaq}(z)$.
Its safe set therefore recovers $57.0\%$ of the true safe set, which is still five times as much as DDH.

This example shows that GP-DDH can construct a useful safe set from data and a choice of kernel alone, without a model or a Lipschitz constant of the system.

\section{Conclusion}
\label{sec:conclusion}

We propose the Gaussian process data-driven Hamiltonian (GP-DDH), a fully data-driven framework that constructs safe sets with probabilistic guarantees under a GP prior on the dynamics.
Its safe sets lie inside the true ones with posterior probability at least $1-\Delta$ for either choice of the confidence parameter.
In the numerical examples, GP-DDH recovered more of the true safe set than DDH while not requiring a model or a Lipschitz constant.
As a future work, we plan to apply GP-DDH to practical systems whose dynamics are hard to model in closed form, such as aircraft involving flight mode transition~\citep{hsu2024towards}.

\subsection*{Generative AI Use}
Generative AI tools (Claude, Anthropic) were used for drafting the manuscript, editing, and writing the implementation and experiment code.
The authors reviewed and verified all AI-assisted content and take responsibility for the final content of the paper.

\acks{This work was supported by the Strategic Hub for International Research Collaboration project of Seoul National University (SNU).}

\bibliography{reference}

\appendix

\section{Proofs of Theorems~\ref{thm:prob_safety} and~\ref{thm:quantile}}
\label{app:proofs}

\subsection{Proof of Theorem~\ref{thm:prob_safety}}

Under Assumption~\ref{asm:vector_valued_gp}, $W$ is a centered Gaussian random field on $\mathcal I$, because each $W(z,p)$ is a fixed linear functional of $f(z)-\mu_0(z)$.
The posterior inherits continuous sample paths from the prior, and the denominator of $W$ is bounded below by $\eta$, so $W$ is continuous on the compact set $\mathcal I$ and hence almost surely bounded given $\mathcal D$.

The Borell--TIS inequality~\citep[Theorem~2.1.1]{adler2007random} then gives $\mathbb E[M(f)\bigm|\mathcal D]<\infty$ and, for every $r>0$,
\begin{equation}
    \label{eq:borell_tis}
    \mathbb P \bigl( M(f) > \mathbb E[M(f)\bigm|\mathcal D] + r \bigm| \mathcal D \bigr) \le \exp \bigl(- r^2/(2\sigma^2) \bigr), \qquad \sigma^2 := \sup_{(z,p) \in \mathcal I} \operatorname{Var}\bigl(W(z,p) \bigm| \mathcal D\bigr).
\end{equation}
Since $\operatorname{Var}(W(z,p)\bigm|\mathcal D)=(p^\top\Sigma_0(z)p)/(p^\top\overline\Sigma(z)p)\le1$, the regularization $\eta^2I_{n_x}$ ensures $\sigma^2\le1$, and taking $r=\sqrt{2\log(1/\Delta)}$ gives \eqref{eq:beta_exact}.

Once \eqref{eq:beta_exact} holds, \eqref{eq:prob_safety} and the safe-set inclusion follow from Lemma~\ref{lem:conservative}, because $M(f)\le\beta$ implies $\gpv\le V$.

\subsection{Proof of Theorem~\ref{thm:quantile}}

Given $\mathcal D$, the random variables $M_h(f),M_{h,1},\ldots,M_{h,S}$ are independent and identically distributed, and hence exchangeable.
The event $M_h(f)>M_{h,(k)}$ occurs if and only if $M_h(f)$ strictly exceeds at least $k$ of the other $S$ values.
Among any $S+1$ real numbers, at most $S+1-k$ of them strictly exceed at least $k$ of the others, so by exchangeability $\mathbb P(M_h(f)>M_{h,(k)}\bigm|\mathcal D)\le(S+1-k)/(S+1)\le\Delta$.
This proves \eqref{eq:quantile_event}.

On the event $\{M_h(f)\le\betaq\}$, $f(z)\in\mathcal F^{\betaq}(z)$ for all $z\in\mathcal Z_h$.
Since $\mathcal Z_h\supseteq\mathcal X_h\times\mathcal U_h$, the argument in the proof of Lemma~\ref{lem:conservative} applied to $(x,u)\in\mathcal X_h\times\mathcal U_h$ yields
\[
    \hat H_h^{\betaq}(x,p)\le\max_{u\in\mathcal U_h}p^\top f(x,u)\le H(x,p)
\]
for all $x\in\mathcal X_h$ and $p\in\mathbb R^{n_x}$.

A monotone scheme preserves this ordering.
Suppose that $\gpv_h\le V_h$ at every node at some time step.
The update with $\gph_h$ applied to $\gpv_h$ is at most the update with $H$ applied to $\gpv_h$, because the update is nondecreasing in the Hamiltonian.
The latter is at most the update with $H$ applied to $V_h$, because the scheme is monotone.
Since $\gpv_h$ and $V_h$ share the terminal and boundary data, induction over the time steps gives $\gpv_h\le V_h$ at every time step.
Linear interpolation in time preserves this ordering on $[0,T]$, which gives \eqref{eq:prob_safety_grid} and the inclusion of the safe sets.

\section{Common Numerical Settings}
\label{app:details}

\paragraph{Gaussian process model.}
Both examples use the separable model of Assumption~\ref{asm:vector_valued_gp} with $\eta=10^{-6}$ and take the maximum over $u$ in \eqref{eq:gph} on the control grid of the HJ solver.
DDH uses the $\ell_2$-ball uncertainty set of Section~\ref{sec:ddh}.

\paragraph{Linear algebra.}
All GP computations are performed in double precision, whereas the precomputed posterior tables and the HJ solver use single precision.
Each Cholesky factorization is first attempted without regularization and, if it fails, retried once with a diagonal jitter $\epsilon=10^{-12}$.
Posterior sample paths are drawn jointly on $\mathcal Z_h$ with the full posterior covariance, since drawing grid points or output components independently would change the distribution of the maximum.
The Mahalanobis norm in \eqref{eq:W} is evaluated by a Cholesky solve rather than by forming $\overline\Sigma(z)^{-1}$ explicitly.

\paragraph{Borell--TIS confidence parameter.}
Since the expectation in \eqref{eq:beta_bt_def} has no closed form, we approximate it from the sample paths $f^{(1)},\ldots,f^{(S)}$ of Section~\ref{sec:order_statistic}.
Applying \eqref{eq:borell_tis} to $W$ restricted to $\mathcal Z_h\times\mathbb S^{n_x-1}$ gives $\mathbb P(M_h(f)\le\mathbb E[M_h(f)\bigm|\mathcal D]+\sqrt{2\log(1/\Delta)}\bigm|\mathcal D)\ge1-\Delta$ with $\mathbb E[M_h(f)\bigm|\mathcal D]\le\mathbb E[M(f)\bigm|\mathcal D]$, so the bound in \eqref{eq:beta_exact} holds regardless of the grid resolution.
Replacing the expectation in \eqref{eq:beta_bt_def} by the sample mean of the $M_{h,s}$ gives the value
\begin{equation}
    \label{eq:beta_bt}
    \frac1S\sum_{s=1}^SM_{h,s}+\sqrt{2\log(1/\Delta)},
\end{equation}
which approximates $\mathbb E[M_h(f)\bigm|\mathcal D]+\sqrt{2\log(1/\Delta)}$ and is used as $\betabt$ in the experiments.

\paragraph{HJ solver.}
The posterior mean and covariance are precomputed at every node of the state grid for each value of the control grid, and the HJ solver reads them directly without interpolation.
The HJ-VIs are solved with a first-order Lax--Friedrichs scheme: the spatial gradient is the average of the forward and backward differences, the dissipation term is $\tfrac12\alpha_i$ times their difference in each coordinate, time stepping is forward Euler, and the obstacle terms are enforced after each step.
At the boundary of the computational domain, one-sided differences are used, so the dissipation vanishes there.

\section{Details of the GP-Sampled Experiment}
\label{app:gp_details}

\paragraph{Problem data.}
The true dynamics \eqref{eq:gp_sampled_oscillator} are defined on $x\in\mathcal X=[-4,4]^2$ and $u\in\mathcal U=[-1,1]$, and $k_{\mathrm{RBF}}$ has unit variance and ARD lengthscales $(0.3,0.3,0.8)$ for $(x_1,x_2,u)$.
The \textit{Avoid} problem uses $g(x)=1-|x_1|$, so that $\unsafeset=\{x:|x_1|>1\}$, and $T=1$.

\paragraph{Data and true dynamics.}
The data set $\mathcal D$ consists of $N=100$ noise-free observations at the first $100$ points of a scrambled Halton sequence on $[-2,2]^2\times\mathcal U$.
The function $f_{\mathrm{true}}$ is drawn jointly on the union of the $15{,}129$ grid pairs and the $100$ design points and is evaluated only at these points, so the data and the true Hamiltonian come from the same draw.
The Cholesky factorization of this prior covariance needed the jitter retry; no other factorization in this experiment did.
The displayed $f_{\mathrm{true}}$ is the third of three independent draws, chosen by inspecting their safe sets for visual clarity.

\paragraph{Grids and confidence parameters.}
The state grid is $41\times41$ on $[-4,4]^2$ (spacing $0.2$), and the control grid consists of $9$ points on $[-1,1]$.
The set $\mathcal Z_h$ is the full product of these grids ($15{,}129$ points), and its posterior covariance admits an exact Cholesky factorization.
The confidence parameters use $\Delta\in\{0.5,0.1,0.01,0.001\}$, and $\betaq$ and $\betabt$ are computed from $S=1{,}000$ paths, so that $k=1{,}000$ for $\Delta=0.001$.

\paragraph{Empirical probability.}
The empirical probabilities in Table~\ref{tab:gp_sampled_results} use $100$ independent repetitions.
Each repetition draws $1{,}000$ calibration paths and $10{,}000$ further paths from the same joint posterior over $\mathcal Z_h$ and recomputes $\betaq$ and $\betabt$ from its own calibration paths.
The columns $\beta$ and Recovery of the table refer to the single set of calibration paths used for Figure~\ref{fig:gp_safe_sets}.
These paths and the $10{,}000$ paths of Figure~\ref{fig:gp_sampled_supM_distribution} come from random streams independent of that set.

\paragraph{Lipschitz constant.}
The constant $L_x=2.42$ is the largest spectral norm of the state Jacobian of $f_{\mathrm{true}}$ over all state and control grid nodes on $[-4,4]^2$, with the Jacobian obtained by second-order finite differences; it is therefore an estimate on the grid rather than a certified bound.

\paragraph{HJ solver.}
The time step is $2\times10^{-4}$ ($5{,}000$ steps to $T=1$), and all methods use the same coefficients $\alpha_i:=\max\{|f_{\mathrm{true},i}|,\ |\mu_{0,i}|+\beta_{\max}\overline\Sigma_{ii}^{1/2}\}$, maximized over all grid pairs, where $\beta_{\max}=8.02$ is the largest $\beta$ in Table~\ref{tab:gp_sampled_results}; this gives $\alpha=(1.27,1.15)$.
Since $|\partial\hat H^\beta_h/\partial p_i|\le|\mu_{0,i}|+\beta\,\overline\Sigma_{ii}^{1/2}$ by the Cauchy--Schwarz inequality, these coefficients bound $|\partial H/\partial p_i|$ for the true and all GP-DDH Hamiltonians on the whole domain, so the scheme is monotone for these methods.
For DDH, whose Hamiltonian varies faster in $p$ far from the data, the bound fails at $7.0\%$ of the states in the reporting window and at $72.9\%$ of all states, almost all of them outside the window, where there are no data.
Enlarging $\alpha$ to cover DDH would smear the true safe set as well, and the computed DDH set nonetheless contains no unsafe state.

\paragraph{Reporting window.}
Results are reported on the window $[-2,2]^2$ ($21\times21$ nodes), and grid states with $|x_1|=1$, where $g=0$, are counted as outside every set, including the true safe set.
The largest characteristic speed multiplied by the horizon ($1.57$) is smaller than the distance between the reporting window and the domain boundary ($2.0$), so the boundary of the computational domain does not affect the reported sets.

\section{Details of the Random Polynomial Experiment}
\label{app:poly_details}

\paragraph{Problem data.}
The dynamics \eqref{eq:2d_poly_dynamics} are defined on $\mathcal U=[-1,1]^2$.
The \textit{ReachAvoid} problem uses $T=0.5$, $l(x) = 0.2^2 - x_1^2 - x_2^2$, and $g(x) = \min\{1-|x_1|, 1-|x_2|\}$, so that $\targetset$ is the disk of radius $0.2$ and $\unsafeset$ is the complement of the closed unit square $[-1,1]^2$.

\paragraph{Data.}
The data set $\mathcal D$ consists of $N=50$ noise-free observations with states and controls drawn uniformly from $[-1,1]^2\times\mathcal U$.

\paragraph{Kernels and hyperparameters.}
The RBF and Mat\'ern-5/2 kernels have ARD lengthscales, and the linear kernel shares the lengthscales of the RBF kernel.
The lengthscales, kernel weights, and $B$ are fitted by Adam with learning rate $0.05$ for $300$ iterations.
No Cholesky factorization of the Gram matrix needed the jitter retry.

\paragraph{Grids and confidence parameters.}
The state grid is $500\times500$ on $[-1.2,1.2]^2$, and the control grid consists of $20\times20$ points on $[-1,1]^2$.
The set $\mathcal Z_h$ consists of $10{,}000$ state--control pairs selected at random without replacement from the $10^8$ grid pairs, and the confidence parameters use $\Delta=0.1$ and $S=1{,}000$.
For Mat\'ern-5/2, the posterior covariance on $\mathcal Z_h$ admits an exact Cholesky factorization.
For RBF and RBF+Lin, it is numerically rank-deficient, and the paths are drawn with a truncated eigendecomposition that discards eigenvalues below $10^{-12}$, which retains $173$ and $182$ of the $10{,}000$ eigenvalues, with a relative Frobenius error below $10^{-7}$.

\paragraph{Lipschitz constant.}
The state Jacobian of \eqref{eq:2d_poly_dynamics} is $[q(u)\ r(u)]$, and $L_x=6.63$ is the maximum of its Frobenius norm over $\mathcal U$, computed by differential evolution, which upper-bounds the spectral norm.

\paragraph{HJ solver.}
The true Hamiltonian is maximized analytically over $\mathcal U$, whereas GP-DDH uses the control grid.
The time step is $10^{-4}$, and all methods use the dissipation coefficients of the true solution.

\end{document}